# On the Electromagnetic Mass Dilemma

Qasem Exirifard, Alessio D'Errico, and Ebrahim Karimi

Nexus for Quantum Technologies, University of Ottawa, 25 Templeton St., Ottawa, Ontario, K1N 6N5 Canada

**We show that a charged sphere moving at a constant velocity $v$ exhibits a mass due to electromagnetic radiation, expressed as $\frac{4}{3+\left(\frac{v}{c}\right)^2}\left(\frac{\mathcal{E}}{c^2}\right)$, where $\mathcal{E}$ is the electromagnetic energy and $c$ the speed of light in vacuum. Our finding reconciles the longstanding mismatch between the electromagnetic mass calculated from the classical electrodynamics' $\frac{4}{3}\left(\frac{\mathcal{E}}{c^2}\right)$ and the relativistic theory.**

In the early formation of electromagnetism, J.J. Thomson proposed the concept of the electron as a physical entity and made initial attempts to determine its mass [1] . Max Abraham [2] and Hendrik Lorentz [3] later advanced this theory, and introduced the idea that the electron's charge is distributed either uniformly throughout a rigid sphere's volume or solely on its surface. This model, known as the Abraham-Lorentz model for the electron, forms a crucial part of our understanding of electromagnetism. When classical radiation calculations are applied to this model, an electromagnetic momentum is assigned that is proportional to the charged sphere's velocity, resulting in what is known as the electromagnetic (EM) "mass" coefficient. The EM mass calculated from classical radiation theory is $m = \left(\frac{4}{3}\right)\frac{\mathcal{E}}{c^2}$, where $\mathcal{E}$ is the EM energy and $c$ is the speed of light in vacuum.

However, the emergence of Einstein's special relativity theory brought a new perspective. According to this theory, a mass, explicitly an inertia mass, can be assigned to fields' energy through the relation of $\mathcal{E} = m\, c^2$. This theory, introduced by Einstein [4], played a significant role in identifying discrepancies between the Abraham-Lorentz model and classical radiation theory. Max Born [5] and Max von Laue [6] consistently observed a mismatch factor of $\left(\frac{4}{3}\right)$, indicating a discrepancy. Enrico Fermi [7] attributed this to the rigid body assumption in the theoretical frameworks of Born and Laue, suggesting that this approximation significantly affects the accuracy of electromagnetic mass calculations.

Henry Poincaré [8] introduced stresses of a non-electromagnetic nature to account for the stability of the Abraham-Lorentz model for electrons. Poincaré added these stresses as a "rubber" force that firmly holds the charges on the sphere and prevents them from departing. Many scientists have tried to find a resolution to this discrepancy. Later, in the 1970s, Richard Feynman [9] pointed out that the discrepancy should disappear once Poincaré stresses are considered. Fritz Rohrlich [10] and Julian Schwinger [11] argued that the Abraham-Lorentz definition of energy and momentum are not relativistically invariant, and they should be altered to something invariant to resolve the discrepancy. R. Becker [12] and V. B. Morozov [13] attributed the discrepancy to the potential elastic energy within the electron.

Let us now briefly review the discrepancy between the EM "mass" calculated by classical electrodynamics and the one of Einstein's special relativity. Consider a charged sphere with a radius, $a$, where its charge, $e$, was distributed on its surfaces uniformly. The electromagnetic energy, the self-energy, associated with this sphere is $\mathcal{E} = e^2/(8\pi\epsilon_0 a) - \epsilon_0$ is the vacuum permittivity. As this charged sphere moves through the free space at a constant velocity $v$, the motion of the charge generates an electric current along its path, inducing an azimuthal magnetic field that accompanies the sphere's radial electric field. The sphere's shape and the 'radial' electric field at high velocities, close to the speed of light, will be deformed. However, they can be analytically calculated precisely via Lorentz boosts. The electromagnetic fields for both cases, when the charged sphere is at rest and when it is moving with a constant velocity of $v$, are shown in Figure 1. The linear momentum of the field can be calculated from the Poynting vector, $\boldsymbol{p} = \epsilon_0 \int dV\,(\boldsymbol{E} \times \boldsymbol{B}) = \frac{1}{6\pi\epsilon_0}\left(\frac{e^2}{ac^2}\right)\boldsymbol{v}$. Therefore, the mass associated with this charged sphere is given by $m_{\text{EM}} = \frac{1}{6\pi\epsilon_0}\left(\frac{e^2}{ac^2}\right)$, and

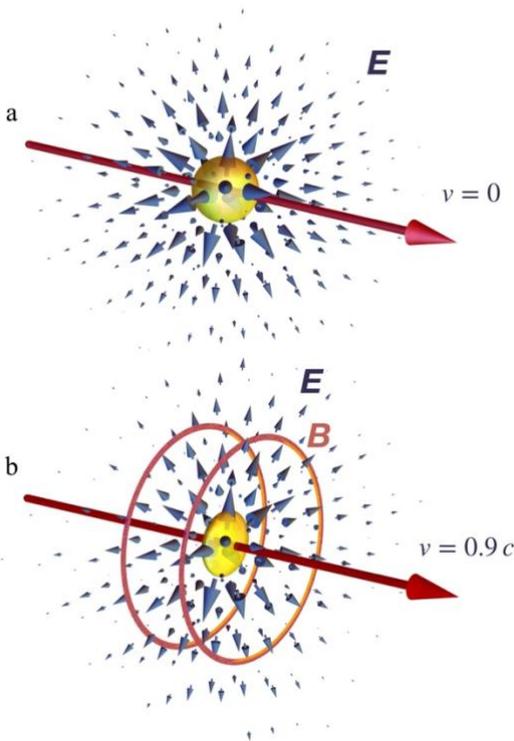

Electric $\boldsymbol{E}$ and magnetic $\boldsymbol{B}$ fields associated with a charged sphere; **a** when it is at the rest, $v = 0$, and **b** when it is moving with a velocity of $v = 0.9\,c$.

thus it can be rewritten as $m_{\text{EM}} = \frac{4}{3}\frac{\mathcal{E}}{c^2}$. This model, of course, assigns a radius to the electron since the self-energy diverges when $a \to 0$; such a radius is known as the classical radius for the electron. The EM mass calculated from the classical electrodynamics differs from what is expected from the famous $\mathcal{E} = mc^2$, i.e. $m = \frac{\mathcal{E}}{c^2}$, where $\mathcal{E}$ is the total energy of the EM field.

In the relativistic regime, objects and fields, and their physical features, follow relativistic transformations, via proper Lorentz boosts $\Lambda^\mu_\nu$, for different observers. Though energy $\frac{\mathcal{E}}{c}$ and momentum $p^i$ denote the total energy and linear momentum of the EM field continuum, they do not compose a four-vector. Within the framework of special relativity, four-vectors are mathematical entities that transform in a specific way under Lorentz transformations. In contrast, the quantities $\left(\frac{\mathcal{E}}{c}, p^i\right)$, defined as spatial integrals over their respective energy and momentum densities—represented by $T^{00}$ and $T^{0i}$—do not follow Lorentz transformations in the same manner as four-vectors. Instead, the total energy $\frac{\mathcal{E}}{c}$ and total momentum $p^i$ of a system are frame-dependent quantities, signifying their reliance on the particular frame of reference from which they are evaluated. Yet, despite their frame-dependency, both the total energy and momentum remain conserved within any given frame. However, unlike the four-vector components, $\left(\frac{\mathcal{E}}{c}\right)^2 - p^2$ is not Lorentz invariant, which marks a fundamental difference between these quantities and the energy-momentum four-vector in the context of special relativity – recall that the energy-momentum four-vector is defined for a point-like objects.

In the context of electrodynamics, within flat spacetime, the electromagnetic energy-momentum tensor $T_{\mu\nu}$ can be calculated from the electromagnetic field, $T_{\mu\nu} = (F_{\mu\lambda}F_\nu^\lambda - \frac{1}{4}g_{\mu\nu}F_{\lambda\sigma}F^{\lambda\sigma})/\mu_0$. Here, $F_{\mu\nu} = \partial_\mu A_\nu - \partial_\nu A_\mu$ is the electromagnetic field tensor, $A_\nu$ is the four potential, $g_{\mu\nu}$ is the Minkowski metric tensor, and $\mu_0$ is the vacuum permeability. The last term of the energy-momentum tensor $F_{\lambda\sigma}F^{\lambda\sigma}$ is a depiction of the field strength tensor contracted with itself. Being a symmetric rank-two tensor, only ten components of the energy-momentum tensor maintain independence in the four-dimensional spacetime geometry. Each of these individual components holds distinct physical representations:

- $T_{00}$: Represents the energy density and is $T_{00} = \frac{1}{2}\left(\epsilon_0|\boldsymbol{E}|^2 + \frac{|\boldsymbol{B}|^2}{\mu_0}\right)$, where $|\boldsymbol{E}|$ and $|\boldsymbol{B}|$ are the electric and magnetic fields magnitude, respectively.
- $T_{0i}$: Illustrates the components of the energy-momentum flux (density) and are given by $T_{0i} = -\epsilon_0(\boldsymbol{E}\times\boldsymbol{B})_i$, where $\times$ stands for the cross product.
- $T_{ii}$ (with $i = \{1,2,3\}$): Represents the pressure or tension and is given by $T_{ii} = \frac{1}{2}\left(\epsilon_0|\boldsymbol{E}|^2 + \frac{|\boldsymbol{B}|^2}{\mu_0}\right) - \left(\epsilon_0|E_i|^2 + \frac{|B_i|^2}{\mu_0}\right)$.
- $T_{ij}$ (for $i \neq j$): Denotes the shear stress which is given by $T_{ij} = -\left(\epsilon_0 E_i E_j + \frac{B_i B_j}{\mu_0}\right)$.

Now it is time to calculate the energy-momentum tensor for our problem, i.e. a charged sphere with a radius, $a$, where its charge, $e$, was distributed on its surfaces uniformly. Employing Gauss's law, the electric field has only the radial component $\hat{r}$ in the spherical coordinates $r, \theta, \phi$, i.e. $E_r(r) = \frac{q}{4\pi\epsilon_0 r^2}$. After calculation, we find the non-zero components of the energy-momentum tensor to be $T_{00} = \frac{\epsilon_0 E_r(r)^2}{2}$, $T_{11} = \frac{\epsilon_0 E_r(r)^2(\cos^2\theta - \sin^2\theta \cos 2\phi)}{2}$, $T_{22} = \frac{\epsilon_0 E_r(r)^2(\cos^2\theta + \sin^2\theta \cos 2\phi)}{2}$, and $T_{33} = \frac{\epsilon_0 E_r(r)^2(\cos 2\theta)}{2}$. Therefore, the total energy and momentum are respectively given by, $\mathcal{E} = \int dV\, T_{00} = e^2/(2a)$ and $p_i = \int dV\, T_{0i} = 0$. Let us now move to a frame in which the charged sphere is moving with the velocity $\boldsymbol{v} = v\,\hat{\boldsymbol{z}}$, i.e. apply a Lorentz boost in the z-direction, which is given by the following $4\times 4$ matrix: $\Lambda^\mu_\nu = \begin{pmatrix} \gamma & 0 & 0 & \beta\gamma \\ 0 & 1 & 0 & 0 \\ 0 & 0 & 1 & 0 \\ \beta\gamma & 0 & 0 & \gamma \end{pmatrix}$, where $\beta = \frac{v}{c}$ and $\gamma = (1-\beta^2)^{-\frac{1}{2}}$. One can compute the elements of the energy-momentum tensor in the new (boosted) frame using the transformation rule $T'^{\mu\nu} = \Lambda^{\mu'}_\mu \Lambda^{\nu'}_\nu T^{\mu\nu}$. By integrating the time-time and time-space components of the transformed tensor, $T'_{00}$ and $T'_{0i}$, respectively —over all $x^{a'}$—we can obtain the total energy and momentum in the boosted frame, $\mathcal{E}' = \gamma\left(1 + \frac{\beta^2}{3}\right)\mathcal{E}$, $p'_x = p'_y = 0$ and $p'_z = \frac{4\,\mathcal{E}}{3\,c}\beta\gamma$. To derive these equations, we used $\int dV = \int dV'/\gamma$. Notice that $\mathcal{E}$ represents the total energy in the rest frame, as calculated in the previous section. Thus, we can find the ratio of momentum $p'_z$ and energy $\frac{\mathcal{E}'}{c}$, which is $\frac{p'_z}{(\mathcal{E}'/c)} = \frac{4\,\beta}{(3+\beta^2)}$ and consequently, the EM mass can be expressed as $m_{EM} = \frac{4}{3+\beta^2}\left(\frac{\varepsilon'}{c^2}\right)$. There are a couple of important observations. Both the energy and momentum components are altered upon the boosts; however, they are not in the forms that are observed for a point particle, i.e., four-vector under Lorentz transformations. Therefore, for the continuum fields – since both momentum and energy are changed upon the Lorentz boost – the electromagnetic mass depends on the observer. For observers that move with a low velocity (when the charged sphere moves with low velocity), $v \ll c$, the electromagnetic mass is reduced to $m_{EM} = \frac{4}{3}\left(\frac{\varepsilon}{c^2}\right)$. While for observers that move at relatively high velocity, e.g. when the sphere moves at the speed of light, then $m_{EM} = \left(\frac{\varepsilon}{c^2}\right)$. In the intermediate cases, the exact relation for the EM mass is $m_{EM} = \frac{4}{3+\beta^2}\left(\frac{\varepsilon'}{c^2}\right)$. Employing this model for electrons requires a finite, non-zero physical dimension to the electron, and thus finite (non-diverging) EM energy. This classical radius for the electron is proportional to the one obtained from the Compton wavelength, i.e. proportional to $\frac{e^2}{m_{EM}c^2}$. Assuming now that an external force, $\mathcal{F}$, acts on the charged sphere. Therefore, it moves with acceleration $\dot{v}$. Since the sphere possesses a charge, it radiates electromagnetic energy; thus, a back-action force is acting on the sphere.

This EM radiation is calculated for a point particle, assuming the electron to have no physical dimension, using the Larmor formula, $\frac{4}{3}\left(\frac{\varepsilon}{c^2}\right)\dot{v}^2$. The back-action force for this point particle is $\frac{4}{3}\left(\frac{\varepsilon}{c^2}\right)\dot{v}$ [14]14. This calculation has been considered to be valid for an extended object, the case of a charged sphere, at the relativistic regime based on the fact that radiation is a Lorentz invariant quantity. Knowing that energy and momentum do not form a four-vector for an extended object, the back-action force now should be recalculated properly for the charged sphere, where it is expected to be $\frac{4}{3+\beta^2}\left(\frac{\varepsilon'}{c^2}\right)\dot{v}$.

In summary, we have shown that classical electromagnetic theory is fully compatible with the special relativity in the concept of EM mass. However, the main challenge is that the momentum and energy of the continuum are composed of a localised charge and its electromagnetic field, which do not follow the conventional four-vector transformation under Lorentz symmetry. Our study focuses on the mass associated with a localised electromagnetic field, which does not conform to a four-vector transformation. This deviation from standard theory suggests a shift in our understanding of electron behaviour. We postulate that a localised electron can be more accurately represented as a condensate – an extended state comprising of photons and electrons. This model deviates from the traditional point-like conception of an electron, introducing a framework where a quantum electron exhibits non-point-like characteristics when localised.

*Acknowledgement:* The authors would like to thank Professors Robert Boyd, Eliahu Cohen, and Gerd Leuchs for fruitful discussions.

*Competing financial interests:* The authors declare no competing interests.